\documentstyle[fleqn,run2col,epsfig,rotate]{article}

\newlength{\dinwidth}
\newlength{\dinmargin}
\newcommand{\PRL}[3]{Phys.~Rev.\ Lett.\ {\bf #1} ({#3}) {#2}}
\newcommand{\PRD}[3]{Phys.~Rev.\ {\bf D#1} ({#3}) {#2}}
\newcommand{\PLB}[3]{Phys.~Lett.\ {\bf B#1} ({#3}) {#2}}
\newcommand{\Comp}[3]{Comput.~Phys.~Comm.\ {\bf #1} ({#3}) {#2}}
\newcommand{\deta}{\Delta \eta}
\newcommand{\SF}{{\cal S}}
\def\gsim{\,\lower.25ex\hbox{$\scriptstyle\sim$}\kern-1.30ex%
\raise 0.55ex\hbox{$\scriptstyle >$}\,}
\def\lsim{\,\lower.25ex\hbox{$\scriptstyle\sim$}\kern-1.30ex%
\raise 0.55ex\hbox{$\scriptstyle <$}\,}

\newcommand{\alps}{\alpha_s}

\def\pythia{P\scalebox{0.8}{YTHIA }}
\def\herwig{H\scalebox{0.8}{ERWIG }}

\def\C2q{C_2(Q)}
\def\tC2q{$\C2q$}
\def\f2q{f_2(Q)}
\def\tf2q{$\f2q$}

\begin{document}  

%
% Some useful tex commands
%
%\begin{titlepage}

\title{ 
Is BFKL ruled out by the Tevatron gaps between jets data?
} 
\author{B.E.~Cox, J.R. Forshaw\address{Department of Physics and Astronomy, University of Manchester, Manchester, M13 9PL, England.} and L. L\"onnblad\address{Dept.~of Theoretical Physics 2, S\"olvegatan 14A, S-223 62  Lund, Sweden.}}
\begin{abstract}
We have performed a detailed phenomenological investigation of the 
hard colour singlet exchange process observed at the Tevatron in
events that have a large rapidity gap between outgoing jets. 
We include the effects of multiple interactions to obtain a prediction for 
the gap survival factor. Comparing the data on the fraction of gap 
events with the prediction from BFKL pomeron exchange we find agreement
provided that a constant value of $\alps$ is used in the BFKL calculation. 
Moreover, the value of $\alps$ is in line with that extracted from 
measurements made at HERA. 
     
\end{abstract}
\maketitle

\section{Introduction}

Events with large rapidity gaps in the hadronic final state and a large momentum transfer across the gap, characterised by the presence of a hard jet on each side of the gap, have been observed in both $p \bar p$ collisions at the Tevatron \cite{D0,D01,CDF,CDF1} and in $\gamma p$ collisions at HERA \cite{ZEUS,H1}. Such events are unexpected in standard Regge phenomenology since the cross section is predicted to fall as $\sim s^{-\alpha |t|}$, where $\alpha \simeq 0.25$ GeV$^{-2}$, whilst events with $|t| > 1000~{\rm GeV}^2$ are routinely observed at the Tevatron. Clearly some other explanation must be sought. Uniquely in diffractive physics, high-$t$ events are amenable to the use of perturbative QCD since the gap producing mechanism is squeezed to small distances \cite{FS}. Such calculations have been carried out within the leading logarithmic approximation of BFKL \cite{BFKL} by Mueller and Tang \cite{MT}, and it is the aim of this talk to present comparisons of these calculations with the latest data from the Tevatron. The situation is greatly complicated by the possibility that rapidity gaps formed by whatever process can be destroyed by multiple interactions between spectator partons in the colliding hadrons. Detailed comparisons made and conclusions drawn from any dynamic model of high-$t$ rapidity gap formation must therefore include a careful treatment of such physics. In this analysis, we use a model implemented in the \pythia Monte Carlo generator to simulate the effects of multi-parton interactions. 
  
\section{D\O\ data versus the BFKL pomeron}
\begin{figure}[h] 
\centerline{\epsfig{file=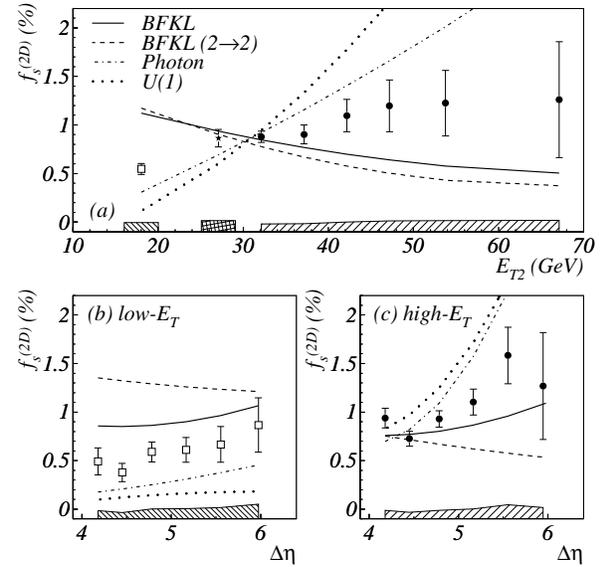,height=7.5cm}}
\caption{D\O\ data compared with a BFKL calculation. Plot from \cite{D01}.}
\label{D0:data}
\end{figure}
The analysis presented here was stimulated to some extent by the recent D\O\ measurements \cite{D01} of the fraction of dijet events containing a large rapidity gap as a function of $E_{T2}$, the $E_T$ of the second hardest jet, and the rapidity difference between the two leading jets, $\Delta \eta$. The D\O\ results are shown in figure \ref{D0:data}. Jets are found using a cone algorithm 
\cite{pxcone,cone} with cone radius $0.7$ and the \verb=OVLIM= 
parameter set to $0.5$. The inclusive dijet sample is defined by the following cuts:
\begin{itemize}
\itemsep 0mm
\item{$|\eta_1|, |\eta_2| > 1.9$, i.e. jets are forward or backward}
\item{$\eta_1 \eta_2 < 0$, i.e. opposite side jets}
\item{$E_{T2} > 15$ GeV}
\item{$\deta > 4$, i.e. jets are far apart in rapidity}.
\end{itemize}
The sub-sample of gap events is obtained by employing the further cut that
there be no particles emitted in the central region $|\eta| < 1$ with
energy greater than 300 MeV. 
 The BFKL curve is clearly ruled out by the data. The D\O\ BFKL curves are based on the calculation of Mueller and Tang implemented into the standard \herwig 5.9 release \cite{herwig,JMB}. In particular, the asymptotic cross-section of \cite{MT} is used; in the limit $y \equiv \deta \gg 1$, 
\begin{equation}
\frac{d \sigma(q q \to q q)}{dt} \approx (C_F \alps)^4 \frac{2\pi^3}{t^2}
\frac{\exp(2 \omega_0 y)}{(7 \alps C_A \zeta(3) y)^3}
\label{BFKL:asym}
\end{equation} 
where $\omega_0 = \omega(0) = C_A (4 \ln 2/\pi) \alps$. The $\alpha_s^4$ in the pre-factor runs with $-t$ according to the two-loop beta function, $\omega_0 = 0.3$ and the $\alpha_s$ in the denominator $=0.25$. The falling of the BFKL curve with increasing $E_{T2}$ is 
driven by the running of the coupling in the pre-factor since the gap 
fraction goes like $\sim \alps^4 / \alps^2$. 

\section{Key issues}

In this analysis, we choose somewhat different parameters. We also use the full Mueller Tang calculation without the asymptotic approximation. This is also available in \herwig 5.9 \cite{MH} and is available from the authors. We choose to fix $\alps = 0.17$. To leading logarithmic accuracy $\alps$ is simply an unknown parameter.
Higher order corrections will indeed cause the coupling to run, however it
is not clear how this should be done in a consistent way. In this paper
we restrict ourselves to the leading logarithmic approximation and treat the
coupling as a free parameter. Moreover, we are guided by recent HERA data
on the double dissociation process \cite{H1:dd} which can be described by the
leading logarithmic BFKL formalism with  
$\alps = 0.17$. We also note that a fixed coupling constant was needed in
order to explain the high-$t$ data on $p\bar{p}$ elastic scattering 
via three gluon exchange \cite{DL}. 
Furthermore, NLO corrections suggest a fixed value for 
the leading eigenvalue of the BFKL equation, $\omega(0)$, \cite{Brodsky} 
which in turn suggests the use of a fixed coupling in the LLA kernel.
\begin{figure}[h] 
\centerline{\epsfig{file=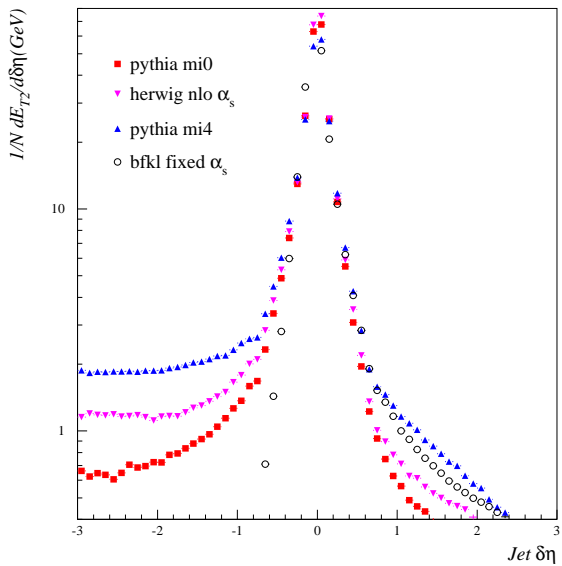,height=8.5cm}}
\caption{Jet $\eta$ profiles}
\label{profiles}
\end{figure}
\subsection*{Underlying events and gap survival}

As mentioned above, it is critical in any estimate of gap formation rates to take into account the possibility that gaps can be destroyed by secondary scatters, which may be perturbative or non-perturbative, between spectator partons in the colliding hadrons. Several models are available \cite{GLM,pythia,jimmy}, but it would be fair to say that all are as yet in a early stage of development and are not tuned to $p \bar p$ data. We choose the model as implemented in \pythia 6.127 \cite{pythia}. Here the
probability to have several parton-parton interactions in the same
collision is modelled using perturbative QCD. The probability for additional interactions is not fixed but varies
according to an impact-parameter picture, where central
collisions are more likely to have multiple interactions. The partons
in the proton are assumed to be distributed according to a double-Gaussian as 
described in \cite{sjo87a,pythia}. There are several parameters in this model 
and we have used the default setting for each.\footnote{Setting
  the switch \texttt{MSTP(82)=4} in \pythia, with everything else
  default, will give the model as we have used it.} Our strategy is to generate high-$t$ photon exchange events (hard BFKL pomeron exchange has not been implemented in \pythia) with and without multiple interactions, and take the percentage change in the number of rapidity gap events, defined as in the D\O\ analysis, as the gap survival factor. We find that gap survival in this model is to first order independent of $E_T^{jet}$ and $\Delta \eta$, i.e. it can be treated as a multiplicative factor. The gap survival factor $\SF$ does vary strongly with centre of mass energy, which is not unexpected since the number density of partons in the colliding hadrons, and therefore the probability of having a secondary scatter, increases with energy. In summary, we find  $\SF(1800~{\rm GeV}) = 22 \%$, 
$\SF(630~{\rm GeV}) = 35 \%$. Full details can be found in \cite{CFL}. 

A key point to notice is the interplay between gap survival and underlying event : multiple interactions also give rise to the so-called jet pedestal and
underlying event effects. This means that the jets measured in
hadron-hadron collisions cannot be compared directly to e.g.\ 
predictions from fixed order perturbation theory. In Figure
\ref{profiles} we show jet profiles obtained from \pythia\ with (mi4) and
without (mi0) multiple interactions (and with $|\delta \phi| < 0.7$). 
The proton remnant is at $\delta \eta > 0$.
It is clear that multiple interactions introduce a jet pedestal of more 
than 1~GeV of $E_T$ per unit rapidity. For comparison, also shown is the jet 
pedestal from \herwig. We note that \herwig\ predicts a greater amount of 
energy outside the jet cone than \pythia\ without multiple interactions. Again, a full discussion of these differences can be found in \cite{CFL}.

In the D\O\ jet measurements the excess $E_T$ from the underlying event
is taken into account by correcting the jet $E_T$ using minimum bias data. 
In particular, the correction
is determined by looking at the $E_T$ flow in regions away from the jets.
The correction is made by subtracting approximately  
1~GeV from the $E_T$ of each reconstructed jet \cite{D0jets}.
In particular, in the gap fraction measurement, this subtraction is
performed for all jets, including those in gap events. But, requiring a large rapidity gap also selects events without 
multiple interactions, where the jet pedestal is absent, or at least much
smaller; multiple interactions destroy gaps, and therefore a gap event {\it cannot have} a multiple interaction. Since jet cross sections fall faster than $1/E_T^4$, such a correction can decrease the measured jet rate by up to 30\% for 18 GeV jets. Our contention therefore is that the jets in gap events should not be corrected for underlying event, and therefore the gap fraction should rise less steeply with $E_T$ than in figure \ref{D0:data}.

\section{Gap fractions}

\begin{figure}[h] 
\centerline{\epsfig{file=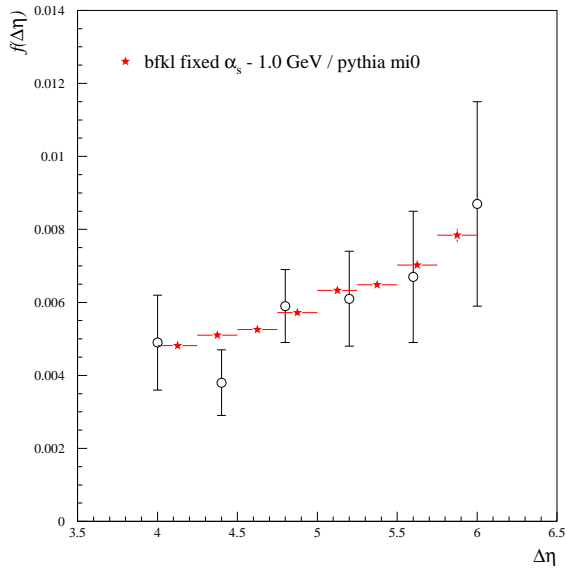,height=8.5cm}}
\caption{Gap fraction as a function of $\Delta \eta$ compared to the D\O\ data}
\label{gf:D01}
\end{figure}

\begin{figure}[h] 
\centerline{\epsfig{file=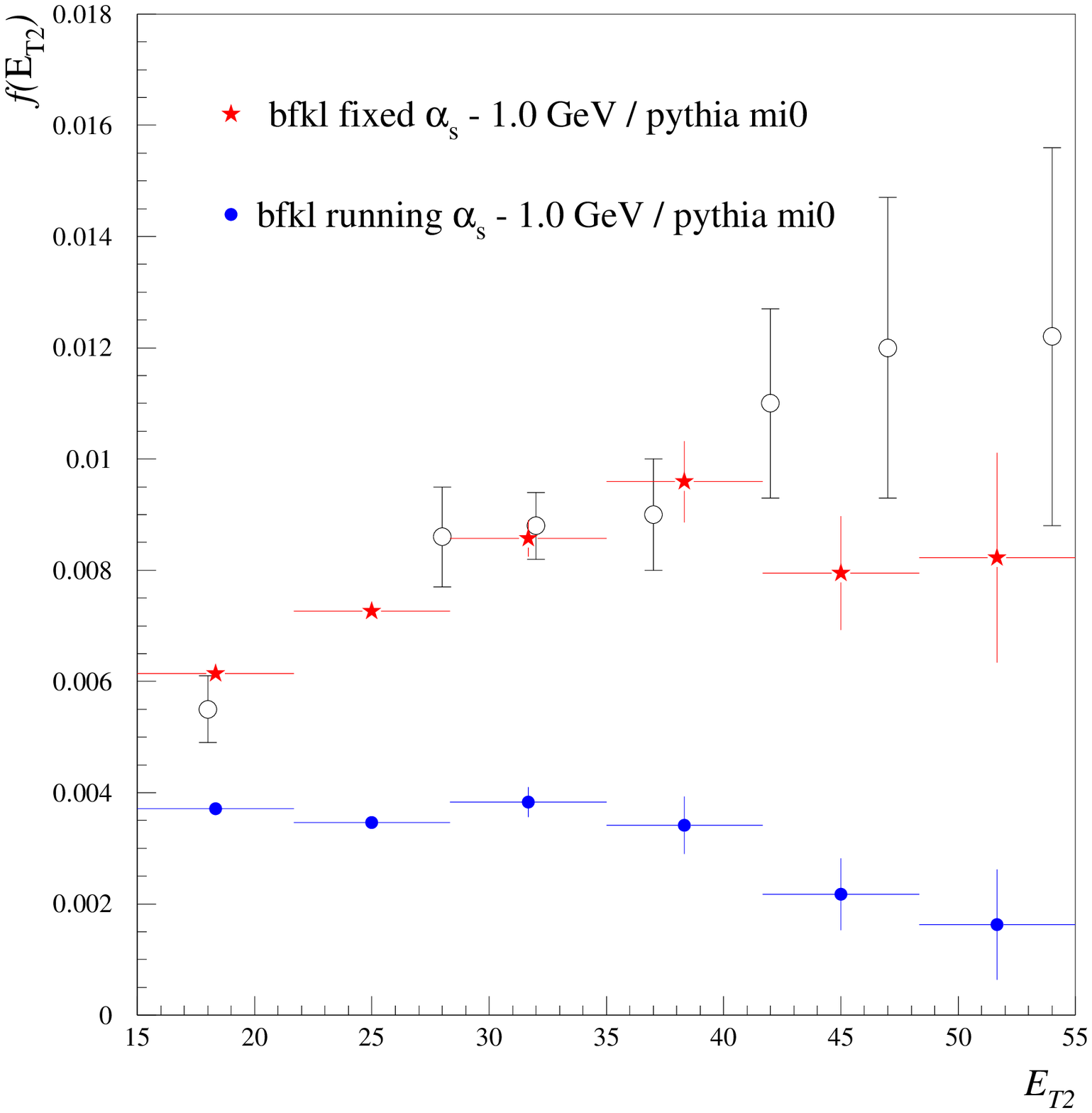,height=8.5cm}}
\caption{Gap fraction as a function of $E_{T2}$ compared to the D\O\ data}
\label{gf:D02}
\end{figure}

\begin{figure}[h] 
\centerline{\epsfig{file=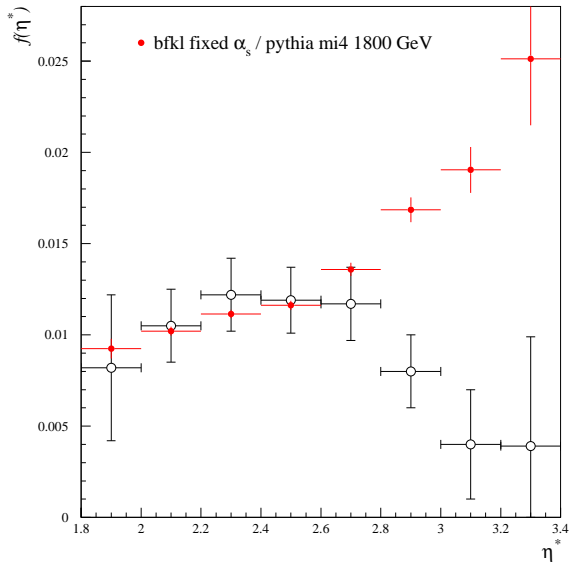,height=8.5cm}}
\caption{Gap fraction compared to the CDF data}
\label{gf:CDF}
\end{figure}

Figures \ref{gf:D01} and \ref{gf:D02} show our results for the gap fractions as functions of $\Delta \eta$ and $E_{T2}$ respectively. The stars are the \herwig BFKL simulation with fixed $\alpha_s = 0.17$, with $1$ GeV subtracted from each jet in order to simulate the D\O\ underlying event correction and the open circles are the D\O\ data. The gap fractions are constructed using a standard \pythia QCD simulation without colour singlet exchange, and without multiple interactions. We have used
both CTEQ2M and CTEQ3M parton distribution functions \cite{CTEQ,PDFLIB}, and 
have found the differences to be small. Our philosophy is that the D\O\ data have been corrected for the effects of multiple interactions in non-singlet exchange events, and we should therefore generate none, whereas we must undo the erroneous correction to the colour singlet sample. The combination of fixing $\alpha_s$ and correcting the gap events erroneously for multiple interactions produces the rise of the gap fraction at low jet $E_T$. The solid circles show the gap fraction using a running $\alpha_s$ in the BFKL sample. Even with the underlying event correction, this sample is unable to fit the data. The overall normalisation of the simulated gap fractions is multiplied by a factor of 0.6. That this is a reasonable thing to do can be appreciated once it is realised that our results have not been fitted to the 
data and that the overall normalisation is acutely sensitive to the magnitude 
of $\alps$. Furthermore, the overall normalisation of the BFKL cross-section 
is uncertain since, within the leading logarithmic approximation, one does not know a priori the scale at which to evaluate the leading logarithms.  Given these points, we conclude 
that the D\O\ data are in agreement with the leading order BFKL result.
Figure \ref{gf:CDF} shows our result for the gap fraction as a function of $\Delta \eta \equiv 2\eta^*$ compared to the CDF data \cite{CDF1}. Note that CDF do not attempt to correct their jets to include the effect of an underlying
event. We therefore generate the \pythia non-singlet sample with multiple interactions (labelled mi4), and do not perform the 1 GeV / jet subtraction from the \herwig BFKL sample. In this plot, our theory points are obtained using a renormalisation 
factor of unity (compared to 0.6 in the D\O\ case). We then find reasonable agreement with the data except at the larger values of 
$\eta^*$ where we are quite unable to explain a fall in the $\eta^*$ 
distribution. Recall however that D\O\ do not see a fall at large $\deta$.
Further clarification of the situation will require an increase in statistics.

We have also computed the ratio of the gap fractions at 630 GeV and 1800 GeV. We find that, even including gap survival effects, $R(630 / 1800) \sim 1$ at the parton level. When hadronisation effects are taken into account, however, we find that the ratio rises significantly to $\sim 3$, with a strong dependence on $\Delta \eta$. D\O\ find $R(630 / 1800) = 3.4 \pm 1.2$ \cite{D01}, and CDF find $R(630 / 1800) = 2.4 \pm 0.9$. In the D\O\ case the effect may be attributed to the different parton $x$ ranges of the 630 GeV and 1800 GeV measurements (although we note that the CDF result is calculated at fixed $x$). The restriction $x<1$ forces the gap and non-gap
cross-sections to fall to zero at some maximum $\deta$, $\deta_{{\rm max}}$.
Now, the colour connection that exists between the jets in the non-gap 
sample drags the jets closer together in rapidity. This has a small effect 
away from $\deta_{{\rm max}}$ (since the $\deta$ spectrum is roughly flat)
however as $\deta \to \deta_{{\rm max}}$ it leads to a more rapid vanishing
of the non-gap cross-section than occurs in the gap cross-section. This
effect, combined with the fact that 
$\deta_{{\rm max}}(630~{\rm GeV}) < \deta_{{\rm max}}(1800~{\rm GeV})$, leads to an enhancement of the measured 630 GeV gap fraction at large $\Delta  \eta$ at the hadron level, and hence the larger value of  $R(630 / 1800)$. 
\section{Conclusions and future possibilities}

We have explicitly demonstrated that the Tevatron data on the gaps-between-jets
process at both 630~GeV and 1800~GeV are in broad agreement with the 
predictions obtained using the leading order BFKL formalism. However, we
are not able to explain the behaviour of the CDF gap fraction at large
$\deta$. Agreement is obtained using the same fixed value of $\alps = 0.17$ 
as was used to explain the recent HERA data on high-$t$ double diffraction 
dissociation. 

Care must be taken in the interpretation of our findings, however. The BFKL formalism itself suffers from being evaluated only to leading logarithmic accuracy. The uncertainties of the overall normalisation which follow will not be removed until an understanding of BFKL dynamics at non-zero $t$ beyond the leading logarithmic approximation in achieved. 

An understanding of the effects of underlying event and its impact on gap survival is crucial to the interpretation of the gaps between jets data, and indeed diffractive data as a whole. 

As pointed out in \cite{MT,CFL}, the gap fraction defined in
terms of a region void of hadronic activity is not strictly infrared safe.
A better observable would be to define a gap to be a region that does not
contain any jets with transverse momenta above some perturbatively large
scale. Work along these lines has also been performed in \cite{OS}. 

One major disadvantage of the gaps between jets process arises from the
need to measure both jets since this limits the reach in rapidity. In
\cite{CF}, it was suggested to focus instead on the double dissociation sample
(the gaps between jets events form a subsample of this generally much
larger sample). By dropping the requirement to observe jets one not
only gains in rapidity reach and statistics but also from the reduced
systematics associated with this more inclusive observable.

\section*{Acknowledgements}
We should like to thank Andrew Brandt, Dino Goulianos, Mark Hayes, Mike 
Seymour and Torbj\"orn Sj\"ostrand for helpful discussions. 
This work was supported by the EU Fourth Framework Programme 
`Training and Mobility of Researchers', Network `Quantum
Chromodynamics and the Deep Structure of Elementary Particles',
contract FMRX-CT98-0194 (DG 12-MIHT). BC would like to thank the UK's
Particle Physics and Astronomy Research Council for support.


\begin{thebibliography}{10}
\itemsep -1mm



\bibitem{D0}
S.~Abachi~et~al~(D\O\ Collaboration), \PRL{72}{2332}{1994};
\PRL{76}{734}{1996}.

\bibitem{D01}
B.~Abbott~et~al~(D\O\ Collaboration), \PLB{81}{189}{1998}.

\bibitem{CDF}
F.~Abe at al (CDF Collaboration), \PRL{74}{855}{1995}; \\
\PRL{80}{1156}{1998}.

\bibitem{CDF1}
F.~Abe at al (CDF Collaboration), \PRL{81}{5278}{1998}. 

\bibitem{ZEUS}
M.~Derrick et al (ZEUS Collaboration), \PLB{369}{55}{1996}. 

\bibitem{H1}
H1 Collaboration, ``Rapidity gaps between jets in Photoproduction at HERA'',
contribution to the International Europhysics Conference on High Energy 
Physics, Jerusalem, Israel (1997).
\bibitem{FS}
J.R.~Forshaw and P.J.~Sutton, Euro.~Phys.~J. {\bf C14} (1998) 285.


\bibitem{BFKL}
I.~Balitsky and L.N.~Lipatov, Sov. J. Nucl. Phys. 28 (1978) 822.

\bibitem{MT}
A.~H.~Mueller and W.~-K.~Tang, \PLB{284}{123}{1992}.

\bibitem{pxcone} S.D.~Ellis, private communication to the OPAL
Collaboration; D.E.~Soper and H.-C.~Yang, private communication 
to the OPAL Collaboration; L.A.~del Pozo, University of Cambridge PhD
thesis, RALT-002 (1993); R.~Akers et al (OPAL Collaboration), Z.~Phys. 
{\bf C63} (1994) 197.

\bibitem{cone} F.~Abe et al (CDF Collaboration), \PRD{45}{1448}{1992}.
\bibitem{herwig}
G.~Marchesini et al, Comp.~Phys.~Comm. {\bf 67} (1992) 465.

\bibitem{H1:dd} 
H1 Collaboration, ``Double Diffraction Dissociation at large $|t|$ in 
Photoproduction at HERA'', contribution to the 29th International Conference 
on High-Energy Physics ICHEP98, Vancouver, Canada, 1998; \\
B.~E.~Cox, ``Double Diffraction Dissociation at large $|t|$ 
from H1'', contribution to the DIS99 Workshop, Zeuthen, Germany (1999) ,hep-ph/9906203.

\bibitem{DL}
A.~Donnachie~and~P.~V.~Landshoff,~Z.~Phys.~{\bf C2} (1979) 55, 
erratum-ibid {\bf C2} (1979) 372; Phys.~Lett. {\bf B387} (1996) 637.

\bibitem{Brodsky}
S.~J.~Brodsky et al, ``The QCD Pomeron with Optimal Renormalisation'', 
SLAC-PUB-8037, IITAP-98-010, hep-ph/9901229.

\bibitem{GLM} E.Gotsman, E.Levin, U.Maor, Phys. Lett. B438 (1998) 229; Phys. Rev. D60 (1999)
\bibitem{pythia}
\pythia\ version 6.127, program and manual,
T.~Sj\"ostrand, \Comp{82}{74}{1994}.
%See also \verb=http://www.thep.lu.se/~torbjorn/Pythia.html=.


\bibitem{jimmy}
J. M. Butterworth, J. R. Forshaw and M. H. Seymour, Zeit. f\"ur Phys. C72 (1996) 637-646. 

\bibitem{sjo87a}
T.~Sj\"ostrand and M.~van Zijl, \PRD{36}{2019}{1987}

\bibitem{CFL} B.E.Cox, J.R. Forshaw, L.L\"onnblad, JHEP10(1999)023.
\bibitem{D0jets}
B.~Abbott et al (D\O\ Collaboration), Nucl.~Instrum.~Methods {\bf A424} 
(1999) 352.
%See also \verb=http://www-d0.fnal.gov/~jkrane/approved_plots=.

\bibitem{JMB}
J.~M.~Butterworth, M.~E.~Hayes, M.~H.~Seymour and L.~E.~Sinclair, ``Rapidity
gaps between jets'', in the proceedings of the Workshop `Future Physics at
HERA', eds. G.~Ingelman, A.~de Roeck and R.~Klanner, DESY (1996).

\bibitem{MH}
M.~E.~Hayes, University of Bristol PhD Thesis (1997).


\bibitem{CTEQ}
H.~L.~Lai et al, \PRD{55}{1280}{1997}.

\bibitem{PDFLIB} 
H.~Plothow-Besch, ``PDFLIB User's Manual'', W5051 PDFLIB, 1997.07.02, 
CERN-PPE; Int.~J.~Mod.~Phys. {\bf A10} (1995) 2901.
\bibitem{OS} G. Oderda and G. Sterman, Phys. Rev. Lett. 81, 3591 (1998).
\bibitem{CF} B. E. Cox and J. R. Forshaw, Phys. Lett. { \bf B434} (1998) 133-140.
\end{thebibliography}
\end{document}